\documentclass[aps,prl,twocolumn]{revtex4}
\usepackage{graphicx}
\usepackage{dcolumn}
\usepackage{amsmath}
\usepackage{amssymb}
\usepackage{subfigure,amsmath,verbatim,moreverb}
\def\rv{{\bf r}}

\begin{document}
\title{Climbing the Density Functional Ladder: Non-Empirical
Meta-Generalized Gradient Approximation Designed for Molecules and Solids}
\author{Jianmin Tao and John P. Perdew}
\affiliation{Department of Physics and Quantum Theory Group, Tulane
University, New Orleans, Louisiana 70118 }
\author{Viktor N. Staroverov and Gustavo E. Scuseria}
\affiliation{Department of Chemistry, Rice University, Houston,
Texas 77005}
\date{\today}
\begin{abstract}
The electron density, its gradient, and the Kohn-Sham orbital
kinetic energy density are the local ingredients of a meta-generalized
gradient approximation (meta-GGA). We construct a meta-GGA density
functional for the exchange-correlation energy that satisfies exact
constraints without empirical parameters. The exchange and
correlation terms respect {\it two} paradigms: one- or two-electron
densities and slowly-varying densities, and so describe both molecules
and solids with high accuracy, as shown by extensive numerical tests.
This functional completes the third rung of ``Jacob's ladder''
of approximations, above
the local spin density and GGA rungs. \par
PACS numbers: 71.15.Mb,31.15.Ew,71.45.Gm
\end{abstract}

\maketitle
Kohn-Sham spin-density functional theory~\cite{KS} reduces
the many-electron ground state problem to a self-consistent
noninteracting-electron form that is exact in principle for the
density and energy, requiring in practice an approximation
for the exchange-correlation
(xc) energy functional $E_{\rm xc}[n_\uparrow,n_\downarrow]$.
No other method achieves comparable accuracy at the same cost.
The exact functional is universal;  its approximations
should be usefully accurate for {\it both} molecules and
solids, and thus for intermediate cases (clusters, biological molecules)
or combinations (chemisorption and catalysis
on a solid surface). The paradigm densities
of quantum chemistry (hydrogen atom and electron pair bond)
and condensed matter physics (uniform electron gas)
thus deserve special respect.
Semi-empirical functionals can fail
outside their fitting sets~\cite{KPB,AFG}; those fitted only
to molecules can be unsuitable for solids.
Alternatively, functionals can be constructed to satisfy
 exact constraints on $E_{\rm xc}[n_\uparrow,n_\downarrow]$.
Non-empirical functionals are not
fitted to actual or computer experiments for real systems,
but are validated by such data.
\par
The first three rungs of ``Jacob's ladder''~\cite{PS} of
approximations can be summarized by the formula
\begin{eqnarray}
 E_{\rm xc}[n_\uparrow,n_\downarrow]  =  \int d^3r \ n\epsilon_{\rm xc}
(n_\uparrow,n_\downarrow,\nabla n_\uparrow,
\nabla n_\downarrow,
\tau_\uparrow,\tau_\downarrow), \
\label{eq_exc}
\end{eqnarray}
where $n(\rv) = n_\uparrow(\rv) + n_\downarrow(\rv)$ is the total
density, and
\begin{eqnarray}
\tau_{\sigma}(\rv) = \sum_{i}^{\mathrm{occup}}\frac{1}{2}
|\nabla\psi_{i \sigma}(\rv)|^2
\label{eq_tau}
\end{eqnarray}
is the kinetic energy density for the occupied Kohn-Sham orbitals
$\psi_{i \sigma}(\rv)$, which are nonlocal functionals of the
density $n_{\sigma}(\rv)$. (We use atomic units where
$\hbar = m = e^2 = 1$.)
The first rung,  found by dropping the
$\nabla n_{\sigma}$ and $\tau_{\sigma}$ dependences in Eq.~(\ref{eq_exc}),
is the local spin density (LSD) approximation~\cite{KS,PW92}, exact
for the uniform electron gas and often usefully
accurate for solids. The second rung, found by dropping only the $\tau_{\sigma}$
dependence in Eq.~(\ref{eq_exc}), is the generalized gradient
approximation (GGA)~\cite{PW86,PBW,PBE} (useful  for molecules as well). The
Perdew-Burke-Ernzerhof (PBE) GGA has {\it two} non-empirical derivations
based on  exact properties of the xc hole~\cite{PBW} and
 energy~\cite{PBE}. \par
The third rung of the ladder, the meta-GGA, makes use of all the
ingredients shown in Eq.~(\ref{eq_exc}). The kinetic energy density
$\tau_{\sigma}(\rv)$ recognizes~\cite{B98} when $n_{\sigma}(\rv)$
has one-electron character by the condition
$\tau_{\sigma}(\rv) = \tau_{\sigma}^{W}(\rv)$, where
$\tau_{\sigma}^{W}(\rv) = |\nabla n_{\sigma}|^2/8n_{\sigma}$
is the von Weizs\"acker kinetic energy density for real orbitals. 
Moreover, Eq.~(\ref{eq_exc}) can
be correct to fourth-order in $\nabla$
for a slowly-varying density~\cite{PKZB}.
However, prior meta-GGAs~\cite{KPB,PKZB,VS98} have been
at least partly empirical, and have not taken full advantage of
$\tau_{\sigma}(\rv)$. The aim of this work is to construct a reliable
non-empirical meta-GGA to complete the third rung of Jacob's ladder. \par
Our starting point is the Perdew-Kurth-Zupan-Blaha (PKZB)
meta-GGA~\cite{PKZB}, which by design yields the correct
exchange and correlation energies through second-order in $\nabla$ for a
slowly-varying density, and the correct correlation energy for any
one-electron density. PKZB has one empirical parameter
(fitted to atomization energies) in its exchange
part, and has demonstrated unexpected successes and failures including:
(i) an accurate strong-interaction limit for the correlation energy of
a spin-unpolarized non-uniform density~\cite{SPK}, (ii) accurate
surface~\cite{KPB,PKZB} and atomization ~\cite{KPB,PKZB,AES} energies,
(iii) poor equilibrium bond
lengths~\cite{AES}, and (iv) poor description of hydrogen-bonded
complexes~\cite{RS}. While PKZB correlation requires only minor
technical improvements, PKZB exchange (the root of the bond-length
errors) needs to take into account
{\it both} paradigm densities. Since it is impossible within the meta-GGA
form to achieve the exact exchange energy for an arbitrary one-
or two-electron density (i.e., the fully nonlocal self-interaction
correction to the Hartree energy), we shall instead require that
the meta-GGA exchange potential be finite at the nucleus for
ground-state one- and two-electron densities, an exact constraint satisfied
by LSD but lost in GGA~\cite{PW86}. This is the key idea of our
construction. Because the PKZB meta-GGA is explained in detail
in Ref.~\cite{PKZB}, we focus here on the added features of our
proposed ``TPSS'' meta-GGA functional. \par
We begin with the exchange energy. Because of the exact
spin-scaling relation~\cite{PKZB}
$E_{\rm x}[n_\uparrow,n_\downarrow] =
E_{\rm x}[2n_\uparrow]/2 + E_{\rm x}[2n_\downarrow]/2$,
where $E_{\rm x}[n] \equiv E_{\rm x}[n/2,n/2]$, we need consider only the
spin-unpolarized case. Using the uniform density-scaling
constraint (Eq. (6) of Ref.~\cite{PKZB}), the meta-GGA (MGGA)
can be written as
\begin{eqnarray}
E_{\rm x}^{\rm {MGGA}}[n] = \int d^3r\ n \epsilon_{\rm x}^{\rm {unif}}(n) F_{\rm x}(p,z),
\label{eq_fxpz}
\end{eqnarray}
where $\epsilon_{\rm x}^{\mathrm{unif}}(n) = -\frac{3}{4\pi}(3\pi^2n)^{1/3}$ is
the exchange energy per particle of a uniform electron gas.
The enhancement factor $F_{\rm x}$ (which equals 1 in LSD)
is a function of two dimensionless
inhomogeneity parameters
\begin{eqnarray}
p = |\nabla n|^2/[4(3\pi^2)^{2/3}n^{8/3}] = s^2
\label{eq_ps}
\end{eqnarray}
and
$z = \tau^{W}/\tau \le 1$,
where $\tau = \sum_\sigma \tau_{\sigma}$ and
$\tau^{W} = \frac{1}{8}|\nabla n|^2/n$. To ensure the Lieb-Oxford
bound (Eq. (7) of Ref.~\cite{PKZB}), we choose
\begin{eqnarray}
F_{\rm x} = 1 + \kappa - \kappa/(1 + x/\kappa),
\label{eq_enhan}
\end{eqnarray}
where $\kappa = 0.804$ and $x(p,z) \ge 0$ will be defined later. 
(A tighter bound~\cite{GH} would make $\kappa = 0.758$.) \par
For a density that varies slowly over space, $F_{\rm x}$ should recover the
fourth-order gradient expansion of Svendsen and von Barth~\cite{SB},
\begin{eqnarray}
F_{\rm x} = 1 + \frac{10}{81}p + \frac{146}{2025} q^2
- \frac{73}{405} q p + D p^2 + O(\nabla^6), \
\label{eq_expan}
\end{eqnarray}
where
$q = \nabla^2 n/[4(3\pi^2)^{2/3}n^{5/3}]$ is the reduced Laplacian.
In PKZB, the gradient coefficient $D$ is an empirical parameter equal
to $0.113$, but in TPSS we set D = 0, the best numerical estimate for
this coefficient~\cite{SB}. The second-order gradient expansion
for $\tau$ (Eq. (4) of Ref.~\cite{PKZB}) makes it possible to
recover Eq.~(\ref{eq_expan}) from a Taylor expansion of
$F_{\rm x}(p,z)$. Note in particular that
\begin{eqnarray}
\tilde{q}_b = (9/20)(\alpha - 1)/
 [1 + b\alpha(\alpha - 1) ]^{1/2} +  2p/3,
\label{eq_dlap}
\end{eqnarray}
an inhomogeneity parameter constructed from $p$ and $z$,
tends to the reduced Laplacian $q$ in the slowly-varying limit.
In Eq.~(\ref{eq_dlap}),
\begin{eqnarray}
\alpha = (\tau - \tau^{W})/\tau^{\mathrm{unif}} = 
(5p/3)(z^{-1} - 1) \ge 0,
\label{eq_alp}
\end{eqnarray}
where $\tau^{\mathrm{unif}} = \frac{3}{10}(3\pi^2)^{2/3}n^{5/3}$
is the uniform-gas kinetic energy density. When the parameter $b$
in Eq.~(\ref{eq_dlap}) is set to zero, $\tilde{q}_b$ reduces
to the dimensionless variable $\tilde q$ defined in Eq. (12)
of Ref.~\cite{PKZB}; $b$ will be defined later. \par
For a two-electron ground-state density, $z = 1$ ($\alpha = 0$)
and the meta-GGA reduces to GGA form with an enhancement factor
$\tilde F_{\rm x}(s) = F_{\rm x}(p = s^2, z = 1)$. The corresponding
exchange potential
$\tilde \upsilon_{\rm x}(\rv) = \delta \tilde E_{\rm x}/\delta n(\rv)$
has a $\nabla^2 n$ term which diverges at a nucleus unless
its coefficient, proportional to $d \tilde F_{\rm x}/ds$
(Eq. (24) of Ref.~\cite{PW86}), vanishes there.
Since, for a two-electron exponential density,
$s$ at a nucleus is 0.376, we eliminate this
spurious divergence (see Fig. 1 of Ref.~\cite{SSTP}) by requiring that
\begin{eqnarray}
d \tilde F_{\rm x}/ds |_{s = 0.376} = 0.
\label{eq_derivative}
\end{eqnarray} 
\par
All of the above exact constraints imposed on the TPSS
meta-GGA, beyond those imposed on the PBE GGA~\cite{PBW,PBE},
concern small dimensionless density derivatives.
We have no reason to modify the large-$p$ behavior of the
PBE enhancement factor $
F_{\rm x} \sim 1 + \kappa - \kappa^2/(\mu p)$ ($p \rightarrow \infty$), 
where $\mu = 0.21951$. Thus we retain this large-$p$ limit,
which can describe
weak binding~\cite{ZPY}. \par
To satisfy all of the above conditions, we choose $x$ of
Eq.~(\ref{eq_enhan}) to be
\begin{eqnarray}
x & = & \Biggl\{ \left [\frac{10}{81} + c\frac{z^2}{(1 + z^2)^2}\right ]p +
\frac{146}{2025} \tilde{q}_b^2 \nonumber \\
& & \nonumber \\
& & - \frac{73}{405}\tilde{q}_b\sqrt{\frac{1}{2}
\left(\frac{3}{5}z \right)^2 + \frac{1}{2}p^2}
+ \frac{1}{\kappa}\left(\frac{10}{81}\right)^2p^2 \nonumber \\
& & \nonumber \\
& &  + 2\sqrt e \frac{10}{81}\left(\frac{3}{5}z \right)^2 + e\mu p^3 \Biggr\}
\Big/(1 + \sqrt e p)^2,
\label{eq_ptx}
\end{eqnarray}
where the constants $c = 1.59096$ and $e = 1.537$ are chosen
to enforce Eq.~(\ref{eq_derivative}) and to yield the correct
exchange energy ($-0.3125$ hartree) for the exact ground-state
density of the hydrogen atom. The form of Eq.~(\ref{eq_ptx}),
which is not uniquely constrained, is chosen to make 
$F_{\rm x}$ smooth and monotonic as a function
of $s$ for $0 \le \alpha \le 1$. Finally, the parameter $b = 0.40$
in Eq.~(\ref{eq_dlap}) is chosen to have the smallest value that
preserves $F_{\rm x}$ as a monotonically increasing function of $s$
even for fixed $\alpha \gg 1$; this is done for esthetic reasons,
since $b = 0$ produces nearly the same results in our molecular
tests. \par
\begin{figure}
\includegraphics[width=\columnwidth]{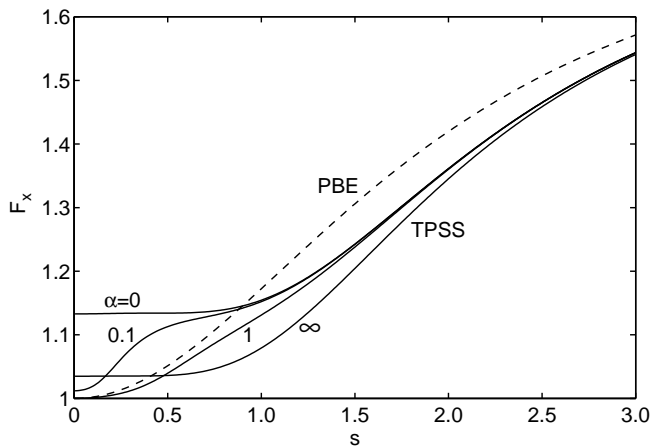}
\caption{TPSS and PBE exchange enhancement factors as
functions of the reduced gradient 
$s$ of Eq.~(\ref{eq_ps}). The TPSS plots (solid lines)
are drawn for four different values of $\alpha$ of 
Eq.~(\ref{eq_alp}). Ground state
two-electron densities have $\alpha = 0$; slowly-varying
densities have $\alpha \approx 1$ and $s \ll 1$.}
\end{figure}
Figure 1 shows that TPSS has a strong but local exchange enhancement
at small $s$ for $\alpha=0$. In contrast, PKZB has $F_{\rm x}\approx 1$
in the small-$s$ region (see the $r_s=0$ curves for $\alpha=0$ and 1
in Figs. 1 and 2 of Ref.~\cite{KPB}), while the PBE GGA $F_{\rm x}$ is altogether
independent of $\alpha$.
   \par
We turn now to  correlation, making minor refinements
(independent of  TPSS exchange) to PKZB~\cite{PKZB}:
\begin{eqnarray}
E_{\rm c}^{\mathrm{MGGA}}[n_\uparrow,n_\downarrow] & = & \int d^3r \ n
\epsilon_{\rm c}^{\mathrm{revPKZB}} \nonumber \\
& &
\times [1 + d \epsilon_{\rm c}^{\mathrm{revPKZB}}
(\tau^{W}/\tau)^3 ],
\label{eq_ectot}
\end{eqnarray}
\begin{eqnarray}
\epsilon_{\rm c}^{\mathrm{revPKZB}} & = & \epsilon_{\rm c}^{\mathrm{PBE}}(n_\uparrow,
n_\downarrow,\nabla n_\uparrow,\nabla n_\downarrow)
 [1 + C(\zeta,\xi) (\tau^{W}/\tau)^2]
\nonumber \\
& &
- [1 + C(\zeta,\xi)](\tau^{W}/\tau)^2
\sum_{\sigma}\frac{n_\sigma}{n} \tilde\epsilon_{\rm c}.
\label{eq_ecden}
\end{eqnarray}
The basic ingredient here is $\epsilon_{\rm c}^{\mathrm{PBE}} 
(n_\uparrow,n_\downarrow,\nabla n_\uparrow,\nabla n_\downarrow)$~\cite{PBE}
which already has the correct second-order gradient
expansion, scales properly under uniform density-scaling to
the high- and low-density (weak- and strong-interaction)
limits, is nonpositive, and vanishes as $p \rightarrow \infty$.
The quantity $\tilde\epsilon_{\rm c}$ could be
$\epsilon_{\rm c}^{\mathrm{PBE}}(n_{\sigma},0,\nabla n_{\sigma},0)$ as in
PKZB, but to ensure the exact constraint $E_{\rm c} \le 0$ for all
possible densities
 we take $\tilde\epsilon_{\rm c} = {\rm max}[
\epsilon_{\rm c}^{\mathrm{PBE}}(n_{\sigma},0,\nabla n_{\sigma},0)$,
$\epsilon_{\rm c}^{\mathrm{PBE}}(n_\uparrow,n_\downarrow,
\nabla n_\uparrow,\nabla n_\downarrow)]$.
The switch is almost seamless  
in the rare cases where it occurs
such as the Li atom (see Fig. 2 of Ref.~\cite{SSTP}), 
and would probably not occur at
all for a perfectly-designed GGA. Note also that
$\tau^{W} = |\nabla n|^2/8n$ was spin-resolved in PKZB.
 \par
For any $C(\zeta,\xi)$, $E_{\rm c}^{\mathrm{MGGA}}$ of Eq.~(\ref{eq_ectot})
properly vanishes for
a one-electron density (for which $\tau = \tau^{W}$ and
$\zeta \equiv (n_\uparrow - n_\downarrow)/n = \pm 1)$.
Since a self-interaction correction should not change the energy
of delocalized electrons, we set $C(0,0) = 0.53$ (as in PKZB)
and $d = 2.8$ hartree$^{-1}$. This choice leaves the surface
correlation energy of jellium unchanged from its PBE GGA values
over the range of valence-electron bulk densities. \par
By properly eliminating the self-correlation error for
spin-unpolarized ($\zeta=0$) densities, the PKZB and TPSS
approximations yield correct correlation energies for atomic
densities (with $\zeta=0$) even in the strong-interaction
limit~\cite{SPK}, where LSD and GGA fail badly. In this limit,
electrons are kept apart by Coulomb repulsion, so
they ``forget" that they are fermions  and
the exchange-correlation energy $E_{\rm xc}$ for a given density
$n(\rv)$  becomes independent of  the relative spin
polarization $\zeta$ (as it almost does in LSD and PBE GGA).
To achieve this independence over the
range $0 \le |\zeta| \le 0.7$ for Gaussian
one-electron and other densities of uniform $\zeta$ requires
\begin{eqnarray}
C(\zeta,0) = 0.53 + 0.87 \zeta^2 + 0.50 \zeta^4 + 2.26 \zeta^6.
\end{eqnarray}
In a monovalent atom like Li, a
self-interaction correction should zero out the correlation
energy density in the valence region without creating any
additional correlation energy density in the core-valence
overlap region (a PKZB error, Fig. 2 of Ref.~\cite{SSTP}). We achieve this  
with
\begin{eqnarray}
\hspace{-0.5cm}
C(\zeta,\xi) = \frac{C(\zeta,0)}{\{1 +  \xi^2
[(1 + \zeta)^{-4/3} + (1 - \zeta)^{-4/3}]/2 \}^4}, \
\label{eq_CCC}
\end{eqnarray}
where $\xi = |\nabla \zeta|/2(3\pi^2n)^{1/3}$ is large in the
overlap region. ($\xi$ also appears in Ref.~\cite{WP}). For
a spin-unpolarized density, setting $d = 0$
in Eq.~(\ref{eq_ectot}) recovers PKZB.
 \par
\begin{table*}
\caption{Statistical summary of the errors of four density functionals
for various properties of molecules, solids, and surfaces.
1 kcal/mol = 0.0434 eV = 0.00159 hartree. For jellium,
$r_s = (3/4\pi n)^{1/3}$ characterizes bulk density.}
\begin{ruledtabular}
\begin{tabular}{llddddddd}
& & \multicolumn{1}{c}{Mean value}
  & \multicolumn{5}{c}{Mean absolute errors}
  & \multicolumn{1}{c}{Mean error} \\ \cline{4-8}
  Property (units) & Test set & \multicolumn{1}{c}{of property}
 & \multicolumn{1}{c}{LSD} & \multicolumn{1}{c}{PBE} &
\multicolumn{1}{c}{PBE0} &
\multicolumn{1}{c}{PKZB} & \multicolumn{1}{c}{TPSS} &
\multicolumn{1}{c}{of TPSS} \\ \hline
Atomization energy $\Sigma D_0$ (kcal/mol) & G2 (148 mols.)
   & 478 & 83.8 & 17.1 & 5.1 & 4.4  & 6.2 & 5.4 \\
Ionization potential (eV) & G2 (86 species)
   & 10.9 & 0.22 & 0.22 & 0.20 & 0.29 & 0.23 & -0.11 \\
Electron affinity (eV) & G2 (58 species)
   & 1.4 & 0.26 & 0.12 & 0.17 & 0.14 & 0.14 & -0.01 \\
Bond length $r_e$ (\AA) & 96 molecules
   & 1.56 & 0.013 & 0.016 & 0.010 & 0.027 & 0.014 & 0.014 \\
Harmonic frequency $\omega_e$ ($\mathrm{cm}^{-1}$) & 82 diatomics
   & 1430 & 48.9 & 42.0 & 43.6 & 51.7 & 30.4 & -18.7 \\
H-bond dissoc. energy $D_0$ (kcal/mol)
   & 10 complexes & 13.4 & 5.8 & 1.0 & 0.7 & 2.9 & 0.6 & 0.2 \\
H-bond lengths $r_e$ (\AA) & 11 H-bonds
   & 2.06 & 0.147 & 0.043 & 0.032 & 0.179 & 0.021 & 0.021 \\
H-bond angles (deg) & 13 angles
   & 111 & 4.0 & 2.6 & 1.8 & 3.5 & 2.0 & 2.0 \\
Lattice constant (\AA) & 17 solids
   & 4.34 & 0.069 & 0.052 & - & 0.078 & 0.037 & 0.035 \\
Bulk modulus (GPa) & 17 solids
   & 124 & 14.6 & 8.7 & - & 9.2 & 9.1 & -1.9 \\
XC surface energy (erg/$\mathrm{cm}^{2}$) & $r_s = 2, 4, 6$
   & 1245 & 22 & 55 & 39 & 5 & 13 & -10 \\
\end{tabular}
\end{ruledtabular}
\end{table*}
Finally, we have made extensive numerical tests of the TPSS
meta-GGA for atoms, molecules, solids and jellium surfaces.
Total energies of atoms are exceptionally accurate, but we
focus in Table I on energy differences and related properties
of greater chemical and physical interest. In this table,
we also report LSD, PBE GGA, PKZB meta-GGA, and PBE0 hybrid~\cite{ES} results for
comparison. Details of our study, and comparisons with other
functionals, will be presented in later publications~\cite{SSTP}. \par
Except for the jellium surface~\cite{KPB}, all our calculations
are performed self-consistently (although post-PBE results would be similar) 
by the method for $\tau$-dependent
functionals described in Ref.~\cite{NNH} using a
developmental version of the Gaussian program~\cite{Gaussian}. 
Alternatively, a meta-GGA Kohn-Sham potential could be found by the
optimized effective potential method~\cite{KP}.  
All molecular calculations 
use the large 6-311++G(3df,3pd) Gaussian basis set. Performance
of the functionals for molecular binding properties is assessed
by computing atomization energies for the 148 molecules
(built up from atoms with $Z \le 17$) of the G2 test set~\cite{CRRP}.
For organic molecules larger than those in this set, TPSS is much
more accurate~\cite{AR,SSTP} than PKZB. 
We also studied the
ten hydrogen-bonded complexes of Ref.~\cite{RS}, reporting
the errors with respect to M\o ller-Plesset (MP2) predictions (which agree
with experimental values where available), and 17 solids
(Li, Na, K, Al, C, Si, SiC, BP, NaCl, NaF, LiCl, LiF, MgO, Cu,
Rh, Pd, Ag in their normal crystal structures) calculated
with various basis sets.
\par
Table I shows that the TPSS meta-GGA gives generally excellent
results for a wide range of systems and properties, 
correcting overestimated PKZB bond lengths in molecules,
hydrogen-bonded complexes, and ionic solids.
We attribute this good performance
to the satisfaction of additional exact constraints on
$E_{\rm xc}[n_\uparrow,n_\downarrow]$, especially  
the key constraint of Eq.~(\ref{eq_derivative}). 
Note also that PBE0~\cite{ES} results are comparable to TPSS, but TPSS
has a practical advantage, since hybrid functionals are not easily
evaluated for solids. More generally, exact exchange is
inaccessible or too slow in many codes.
 \par
The atomization energy of a molecule can be regarded as the
extra surface energy of the separated atoms. Meta-GGAs
are sophisticated enough to reduce the too-large LSD
atomization energies and increase the too-small
LSD jellium surface energies, while GGAs reduce both. \par
While LSD and PBE GGA are controlled extrapolations from the
slowly-varying limit, TPSS meta-GGA is more like a
controlled interpolation between slowly-varying and
one- or two-electron limits.
Our functionals are nested like Chinese boxes:
LSD is inside PBE GGA, and PBE GGA is inside TPSS meta-GGA.
In a future work, we will carry TPSS forward into the fourth
rung of Jacob's ladder, where an exact-exchange ingredient is employed
in a global hybrid~\cite{B98} like PBE0, a local hybrid~\cite{JSE}, or a
hyper-GGA~\cite{PS}. A hyper-GGA is exact for any one-electron
density, and for any density scaled uniformly to the
high-density limit. \\

Acknowledgments: J.T. and J.P.P acknowledge the support of the
National Science Foundation under Grant No. DMR-01-35678 and
discussions with S. K\"ummel and F. Furche.
V.N.S. and G.E.S. acknowledge the support of
the National Science Foundation and Welch Foundation.

\end{document}